\documentclass[aps,prd,twocolumn,superscriptaddress]{revtex4}
\usepackage{epsfig,epsf}
\usepackage{amsmath}
\usepackage{amsthm}
\usepackage{amsfonts}
\usepackage{amssymb}
\usepackage{dsfont}
\usepackage{multirow}
\usepackage{appendix}
\usepackage{slashed}
\usepackage[active]{srcltx}
\usepackage{psfrag}

\begin{document}

\title{Scalar exotic mesons $bb\overline{c}\overline{c}$}
\date{\today}
\author{S.~S.~Agaev}
\affiliation{Institute for Physical Problems, Baku State University, Az--1148 Baku,
Azerbaijan}
\author{K.~Azizi}
\thanks{Corresponding Author}
\affiliation{Department of Physics, University of Tehran, North Karegar Avenue, Tehran
14395-547, Iran}
\affiliation{Department of Physics, Do\v{g}u\c{s} University, Dudullu-\"{U}mraniye, 34775
Istanbul, T\"{u}rkiye}
\author{B.~Barsbay}
\affiliation{Division of Optometry, School of Medical Services and Techniques, Do\v{g}u%
\c{s} University, 34775 Istanbul, T\"{u}rkiye}
\author{H.~Sundu}
\affiliation{Department of Physics Engineering, Istanbul Medeniyet University, 34700
Istanbul, T\"{u}rkiye}

\begin{abstract}
Properties of doubly charged scalar tetraquarks $bb\overline{c}\overline{c}$
are investigated in the framework of the QCD sum rule method. We model them
as diquark-antidiquark states $X_{\mathrm{1}}$ and $X_{\mathrm{2}}$ built of
axial-vector and pseudoscalar diquarks, respectively. The masses and current
couplings of these particles are computed using the QCD two-point sum rule
method. Results $m_{1}=(12715 \pm 80)~\mathrm{MeV}$ and $m_{2}=(13370 \pm
95)~\mathrm{MeV}$ obtained for the masses of these particles are used to
determine their kinematically allowed decay modes. The full width $\Gamma_{%
\mathrm{1}}$ of the state $X_{\mathrm{1}}$ is evaluated by taking into
account its strong decays to mesons $2B_{c}^{-}$, and $2B_{c}^{\ast -}$. The
processes $X_{\mathrm{2}} \to 2B_{c}^{-}$, $2B_{c}^{\ast -}$ and $%
B_{c}^{-}B_{c}^{-}(2S)$ are employed to estimate $\Gamma_{\mathrm{2}}$.
Predictions obtained for the full widths $\Gamma_{\mathrm{1}}=(63 \pm 12)~%
\mathrm{MeV}$ and $\Gamma_{\mathrm{2}}=(79 \pm 14)~\mathrm{MeV}$ of these
structures and their masses may be utilized in experimental studies of fully
heavy resonances.
\end{abstract}

\maketitle


\section{Introduction}

\label{sec:Intro}

During last few years, the fully heavy-flavor four-quark mesons became one
of the hot topics in the high energy physics. The close interest to this
class of uncanonical hadrons is connected with the new experimental
achievements in related areas. Recent discoveries of $X$ structures made by
the LHCb, ATLAS, and CMS Collaborations in di-$J/\psi $ and $J/\psi \psi
^{\prime }$ mass distributions are among such important results \cite%
{LHCb:2020bwg,Bouhova-Thacker:2022vnt,CMS:2023owd}. New resonances labeled $%
X(6200)$, $X(6600)$, $X(6900)$, and $X(7300)$ have the masses in the range $%
6.2\div 7.3~\mathrm{GeV}$ and are supposedly tetraquarks built of valence
charm quarks and antiquarks.

Theoretical studies of fully heavy-flavor tetraquarks have considerably
longer history than their experimental investigations. Here, we restrict
ourselves by considering mainly articles devoted to analysis of the
LHCb-ATLAS-CMS data, and refer to Ref.\ \cite{Agaev:2023wua} for information
about earlier publications.

There are two mainstreams in the literature competing with each other to
interpret $X$ structures. First of them is based on attempts to treat
observed resonances as coupled-channel effects, and calculate their masses
and widths using this picture, i.e., to explain new structures by
interactions of conventional mesons. This approach was activated in Refs.\
\cite{Dong:2020nwy,Liang:2021fzr} to analyze $X$ structures as well. In the
framework of the second paradigm one considers $X$ as tetraquark resonances.
Within the tetraquark paradigm there are also two branches, in which $X$
resonances are examined either as diquark-antidiquark or hadronic
molecule-type systems. It is worth noting that superpositions of these two
structures can also be employed to reach agreements with available data.
Related problems were addressed in numerous publications \cite%
{Zhang:2020xtb,Albuquerque:2020hio,Yang:2020wkh,Becchi:2020mjz,Becchi:2020uvq,Wang:2022xja,Faustov:2022mvs,Niu:2022vqp,Dong:2022sef,Yu:2022lak,Kuang:2023vac,Wang:2023kir}%
.

Production mechanisms of fully charmed (in general, fully heavy) tetraquarks
constitute another field of interesting investigations \cite%
{Berezhnoy:2011xn,Karliner:2016zzc,Feng:2023agq,Abreu:2023wwg}. Observation
of $X$ resonances prove that generation of such structures is already
accessible at present energies of $pp$ collisions. There are different
suggestions about underlying partonic subprocesses that create four $c$
quarks and about their fragmentation to exotic mesons. Some of these
mechanisms seem complement each other and should not be considered as
alternative ones.

The $X$ resonances were studied also in our works \cite%
{Agaev:2023wua,Agaev:2023ruu,Agaev:2023gaq,Agaev:2023rpj}, in which we used
the QCD two- and three-point sum rule methods to calculate spectroscopic
parameters and widths of different fully charmed scalar diquark-antidiquark
and molecule models. In accordance with our results, the resonance $X(6200)$
is presumably the hadronic molecule $\eta _{c}\eta _{c}$ \cite{Agaev:2023ruu}%
, whereas $X(6600)$ can be considered as the diquark-antidiquark state made
of axial-vector components \cite{Agaev:2023wua}. The tetraquark composed of
pseudoscalar diquarks and hadronic molecule $\chi _{c0}\chi _{c0}$ lead to
predictions which are consistent with the mass and full width of the
resonance $X(6900)$ \cite{Agaev:2023ruu,Agaev:2023gaq}. Therefore, this
structure maybe is a superposition of the diquark-antidiquark and
molecule-type states. The last resonance from this list $X(7300)$ may be
interpreted as an admixture of the hadronic molecule $\chi _{c1}\chi _{c1}$
and first radial excitation of $X(6600)$ \cite{Agaev:2023rpj}.

Important goals in theoretical investigations of heavy four-quark mesons
still are structures stable against strong decays. Such tetraquarks can
transform to conventional particles (mesons, leptons) only due to
electro-weak decays. Therefore, they should be very narrow states with long
mean lifetime. Analyses performed in the context of different methods
confirm that some of tetraquarks containing heavy diquark and light
antidiquark are strong-interaction stable particles \cite%
{Carlson:1987hh,Navarra:2007yw,Karliner:2017qjm,Eichten:2017ffp,Agaev:2018khe}%
. The axial-vector tetraquark $T_{bb}^{-}=bb\overline{u}\overline{d}$ with
the mass below the $B^{-}\overline{B}^{\ast 0}$ and $B^{-}\overline{B}%
^{0}\gamma $ thresholds is one of such exotic mesons. The width of this
state was evaluated in Ref.\ \cite{Agaev:2018khe} and found around of $%
10^{-7}~\mathrm{MeV}$. The parameters of $T_{bb}^{-}$ and its weak decays
were also examined in Ref.\ \cite{Hernandez:2019eox}.

It is remarkable, that the axial-vector state $T_{cc}^{+}=cc\overline{u}%
\overline{d}$ containing heavy diquark $cc$ was discovered by LHCb in the
mass distribution of $D^{0}D^{0}\pi ^{+}$ mesons \cite{LHCb:2021auc}. It has
very small width $410~\mathrm{keV}$ and is the longest living tetraquark
seen experimentally. In light of this fact, one may expect that the beauty
partner of $T_{cc}^{+}$, i.e., the tetraquark $T_{bb}^{-}$ is really stable
against strong decays and will be seen experimentally in the near future.

The fully beauty tetraquarks were investigated in numerous publications, for
instance, in Refs.\ \cite%
{Berezhnoy:2011xn,Karliner:2016zzc,Esposito:2018cwh,Anwar:2017toa,Chen:2016jxd}%
, with results contradictory in some aspects. Thus, in Ref.\ \cite%
{Berezhnoy:2011xn} it was demonstrated that the scalar tetraquark $X_{%
\mathrm{4b}}$ cannot decay to $\eta _{b}\eta _{b}$ mesons, whereas in Ref.\
\cite{Karliner:2016zzc} the mass of $X_{\mathrm{4b}}$ was found below $%
\Upsilon \Upsilon $ but higher than $\eta _{b}\eta _{b}$ thresholds. In our
articles \cite{Agaev:2023wua,Agaev:2023gaq}, we computed the parameters of
the fully beauty scalar tetraquarks $X_{\mathrm{4b}}$ and $T_{\mathrm{4b}}$
by modeling them as states built of axial-vector and pseudoscalar diquarks,
respectively. The mass of $X_{\mathrm{4b}}$ was found residing below the $%
\eta _{b}\eta _{b}$ threshold, whereas $T_{\mathrm{4b}}$ is above $\eta
_{b}\eta _{b}$ but below $\Upsilon \Upsilon $ mass limits.

The fully heavy-flavor tetraquarks $cc\overline{c}\overline{c}$ and $bb%
\overline{b}\overline{b}$ cannot be stable against strong decays, because
annihilations of $b\overline{b}$ or $c\overline{c}$ to gluon(s) generate
their decays to a pair of open-flavor heavy mesons \cite%
{Becchi:2020mjz,Becchi:2020uvq}. It is clear that dominant decay channels of
such structures are processes with charmonia or bottomonia pairs at the
final state. These decay modes are responsible for the bulk of their full
widths. But in a situation when the mass of a fully heavy-flavor tetraquark
is less than the corresponding thresholds, we may encounter a relatively
narrow state. In the case of the tetraquarks $cc\overline{c}\overline{c}$
this threshold is determined by the mass of $2\eta _{c}$ mesons, whereas for
structures $bb\overline{b}\overline{b}$ the corresponding limit is the mass
of $2\eta _{b}$ pair. In Ref.\ \cite{Agaev:2023ara}, we evaluated full
widths of $X_{\mathrm{4b}}$ and $T_{\mathrm{4b}}$ through their decays to $%
B_{q}\overline{B}_{q}$ and $B_{q}^{\ast }\overline{B}_{q}^{\ast }$ mesons.
Obtained prediction $9.6~\mathrm{MeV}$ for the full width of $X_{\mathrm{4b}%
} $ proves that it is a relatively narrow state being, nevertheless,
incomparably wider than $T_{bb}^{-}$ .

To find strong-interaction stable fully heavy-flavor tetraquarks, it is
necessary to exclude channels generated by annihilations of heavy
quark-antiquark pairs. This option is realized in structures $bb\overline{c}%
\overline{c}$ which, evidently, may decay to $B_{c}$ meson pairs, but are
stable against strong decays provided their masses are below relevant
thresholds. The tetraquarks $bb\overline{c}\overline{c}$ and their charge
conjugate states $cc\overline{b}\overline{b}$ are double-charged particles,
and in this sense, are interesting as well.

The first double-charged scalar resonance $T_{cs0}^{a}(2900)^{++}$ was seen
recently by the LHCb Collaboration in the $D_{s}^{+}\pi ^{+}$ mass
distribution of the decay $B^{+}\rightarrow D^{-}D_{s}^{+}\pi ^{+}$ \cite%
{LHCb:2022xob,LHCb:2022bkt}. The LHCb also observed its neutral partner $%
T_{cs0}^{a}(2900)^{0}$. These tetraquarks have the quark-contents $cu%
\overline{s}\overline{d}$ and $cd\overline{s}\overline{u}$ and are first
fully open-flavor four-quark systems fixed experimentally. Such structures
were predicted and studied in the diquark-antidiquark model in Refs.\ \cite%
{Agaev:2016lkl,Chen:2016mqt}. Suggestions to search for double-charged
open-flavor tetraquarks were made in Refs.\ \cite{Chen:2017rhl,Agaev:2017oay}%
, in which the authors calculated their spectroscopic parameters and
explored allowed decay channels. Partial widths of some of these modes were
computed as well.

The structures $bb\overline{c}\overline{c}$ also deserve detailed analyses,
because they contain two quark flavors and carry two units of the electric
charge. Moreover, there are hopes that some of these particles may be stable
against strong decays. Properties of the tetraquarks $bb\overline{c}%
\overline{c}$/$cc\overline{b}\overline{b}$ with different spin-parities were
investigated in various articles \cite%
{Wu:2016vtq,Li:2019uch,Wang:2019rdo,Liu:2019zuc,Wang:2021taf}. Conclusions
made in these publications about stability of tetraquarks $bb\overline{c}%
\overline{c}$ are controversial. For example, investigations performed in
Ref.\ \cite{Wu:2016vtq} using the color-magnetic interaction show that the
scalar, axial-vector and tensor particles $bb\overline{c}\overline{c}$ have
masses overshooting the relevant $B_{c}B_{c}$ thresholds. The similar
conclusions about the tetraquarks $bb\overline{c}\overline{c}$/$cc\overline{b%
}\overline{b}$ with quantum numbers $J^{\mathrm{P}}=0^{+}$, $1^{+}$, and $%
2^{+}$ were drawn in Ref.\ \cite{Liu:2019zuc}: It was argued that these
particles are above their lowest open flavor decay channels for about $300~%
\mathrm{MeV}$. Contrary, in Ref.\ \cite{Wang:2021taf} the charge conjugation
structures with the same spin-parities were found below the $B_{c}B_{c}$
mass limits, and hence they can transform to conventional particles only
through radiative transitions or weak decays. The weak semileptonic and
non-leptonic decays of the scalar tetraquark $bb\overline{c}\overline{c}$
were studied in Ref.\ \cite{Li:2019uch}.

Though $bb\overline{c}\overline{c}$/$cc\overline{b}\overline{b}$ tetraquarks
are yet hypothetical particles and have not been discovered experimentally,
analyses demonstrate that a search for these states is feasible in the
future runs of LHC and in Future Circular Collider \cite%
{Feng:2023agq,Abreu:2023wwg}. Therefore, there is a necessity to undergone
such systems to more detailed consideration. In the present article, we
address namely these questions by calculating masses and widths of the
scalar tetraquarks $bb\overline{c}\overline{c}$. We model them as
diquark-antidiquark states $X_{\mathrm{1}}$ and $X_{\mathrm{2}}$ with $%
C\gamma _{\mu }\otimes \gamma ^{\mu }C$ and $C\otimes C$ structures, and
color $\overline{\mathbf{3}_{c}}\otimes \mathbf{3}_{c}{}$ (in brief form,
"triplet") and $\mathbf{6}_{c}\otimes \overline{\mathbf{6}}_{c}$ ("sextet")
organizations, respectively.

This article is structured in the following way: In Sec.\ \ref{sec:Scalar},
we calculate the masses and current couplings of two scalar models with
different inner structures. We explore the allowed decay channels of the
scalar tetraquarks and evaluate the full width of $X_{\mathrm{1}}$ in Sec.\ %
\ref{sec:ScalarWidths1}. To this end, we compute the partial widths of the
processes $X_{\mathrm{1}}\rightarrow B_{c}^{-}B_{c}^{-}$ and $X_{\mathrm{1}%
}\rightarrow B_{c}^{\ast -}B_{c}^{\ast -}$. The similar consideration for
the tetraquark $X_{\mathrm{2}}$ is performed in Sec.\ \ref{sec:ScalarWidths2}%
. Here, besides decays to the $B_{c}^{-}B_{c}^{-}$ and $B_{c}^{\ast
-}B_{c}^{\ast -}$ final states, we consider also the channel $X_{\mathrm{2}%
}\rightarrow B_{c}^{-}B_{c}^{-}(2S)$. We analyze obtained results in Sect.\ %
\ref{sec:Conc} by comparing them with predictions available in the
literature.


\section{Spectroscopic parameters of the scalar tetraquarks}

\label{sec:Scalar}

The masses $m_{1(2)}$ and current couplings $\Lambda _{1(2)}$ of the scalar
tetraquarks $X_{\mathrm{1(2)}}=bb\overline{c}\overline{c}$ are important
parameters necessary to determine their possible decay channels. The pairs
of spectroscopic parameters $m_{1}$, $\Lambda _{1}$ and $m_{2}$, $\Lambda
_{2}$ can be extracted by means of the two-point sum rule (SR) method \cite%
{Shifman:1978bx,Shifman:1978by}-- one of effective techniques to investigate
the ordinary and multiquark hadrons.

To find the sum rules for these quantities, one has to consider the
correlation function%
\begin{equation}
\Pi (p)=i\int d^{4}xe^{ipx}\langle 0|\mathcal{T}\{J(x)J^{\dag
}(0)\}|0\rangle .  \label{eq:CF1}
\end{equation}%
where $J(x)$ is an interpolating current for one of these scalar particles.
The symbol $\mathcal{T}$ \ above indicates the time-ordering of two $J(x)$
and $J^{\dag }(0)$ currents' product.

It is evident that all information about tetraquarks $X_{\mathrm{1(2)}}$ are
encoded in the currents $J_{1}(x)$ and $J_{2}(x)$. We use two models for the
scalar tetraquark. In the first case, we consider $X_{\mathrm{1}}$ as a
state containing the axial-vector diquark $b^{T}C\gamma _{\mu }b$ and
antidiquark $\overline{c}\gamma ^{\mu }C\overline{c}^{T}$. Then, the
interpolating current that corresponds to this structure has the form
\begin{equation}
J_{1}(x)=[b_{a}^{T}(x)C\gamma _{\mu }b_{b}(x)][\overline{c}_{a}(x)\gamma
^{\mu }C\overline{c}_{b}^{T}(x)],  \label{eq:CR1}
\end{equation}%
where $a$, $b$ are color indices and $C$ is the charge conjugation operator.
\ This current is antisymmetric in color indices and belongs to $[\overline{%
\mathbf{3}_{c}}]_{bb}\otimes \lbrack \mathbf{3}_{c}]_{\overline{c}\overline{c%
}\text{ }}$ representation of the $SU_{c}(3)$ color group. In the second
case, we choose a scalar particle built of \ the pseudoscalar diquarks, and
to study it employ the current $J_{2}(x)$
\begin{equation}
J_{2}(x)=[b_{a}^{T}(x)Cb_{b}(x)][\overline{c}_{a}(x)C\overline{c}%
_{b}^{T}(x)],  \label{eq:CR2}
\end{equation}%
which has the color $[\mathbf{6}_{c}]_{bb}\otimes \lbrack \overline{\mathbf{6%
}}_{c}]_{\overline{c}\overline{c}\text{ }}$organization.

In what follows, we write down formulas for the tetraquark $X_{\mathrm{1}}$.
Expressions for the state $X_{\mathrm{2}}$ can be obtained from them
trivially. The physical side of the sum rule $\Pi _{1}^{\mathrm{Phys}}(p)$
is
\begin{equation}
\Pi _{1}^{\mathrm{Phys}}(p)=\frac{\langle 0|J_{1}|X_{\mathrm{1}}\rangle
\langle X_{\mathrm{1}}|J_{1}^{\dagger }|0\rangle }{m_{1}^{2}-p^{2}}+\cdots ,
\label{eq:Phys1}
\end{equation}%
where the term presented explicitly is a contribution to $\Pi _{1}^{\mathrm{%
Phys}}(p)$ coming from the ground-state particle. Here, the dots indicate
effects of higher resonances and continuum states.

It is convenient to write down $\Pi _{1}^{\mathrm{Phys}}(p)$ using the
parameters of the tetraquark $X_{\mathrm{1}}$, i.e., using the mass and
current coupling of this particle. To this end, we employ the following
matrix element
\begin{equation}
\langle 0|J_{1}|X_{\mathrm{1}}\rangle =\Lambda _{1}.  \label{eq:ME1}
\end{equation}%
Then, the expression for $\Pi _{1}^{\mathrm{Phys}}(p)$ takes a simple form%
\begin{equation}
\Pi _{1}^{\mathrm{Phys}}(p)=\frac{\Lambda _{1}^{2}}{m_{1}^{2}-p^{2}}+\cdots .
\label{eq:Phys2}
\end{equation}%
Because the correlation function $\Pi _{1}^{\mathrm{Phys}}(p)$ has the
Lorentz structure proportional to $\mathrm{I}$, the right-hand side of Eq.\ (%
\ref{eq:Phys2}) forms the invariant amplitude $\Pi _{1}^{\mathrm{Phys}%
}(p^{2})$ necessary for further analysis.

The QCD side $\Pi _{1}^{\mathrm{OPE}}(p)$ of the SR must be calculated in
terms of the heavy quark propagators with certain accuracy in the operator
product expansion ($\mathrm{OPE}$). In the case under consideration, we take
into account in $\Pi _{1}^{\mathrm{OPE}}(p)$ the perturbative term and a
nonperturbative contribution which is proportional to $\langle \alpha
_{s}G^{2}/\pi \rangle $. The reason is that, heavy quark propagators do not
depend on light quark and mixed quark-gluon condensates, and next terms in $%
\mathrm{OPE}$ are ones $\sim \langle g_{s}^{3}G^{3}\rangle $ and $\langle
\alpha _{s}G^{2}/\pi \rangle ^{2}$, which can be safely neglected.

Computations of $\Pi _{1}^{\mathrm{OPE}}(p)$ lead to its expression in terms
of the heavy quark propagators
\begin{eqnarray}
&&\Pi _{1}^{\mathrm{OPE}}(p)=i\int d^{4}xe^{ipx}\left\{ \mathrm{Tr}\left[
\gamma _{\mu }\widetilde{S}_{c}^{a^{\prime }b}(-x)\gamma _{\nu
}S_{c}^{b^{\prime }a}(-x)\right] \right.  \notag \\
&&\times \left[ \mathrm{Tr}\left[ \gamma ^{\nu }\widetilde{S}%
_{b}^{ba^{\prime }}(x)\gamma ^{\mu }S_{b}^{ab^{\prime }}(x)\right] -\mathrm{%
Tr}\left[ \gamma ^{\nu }\widetilde{S}_{b}^{aa^{\prime }}(x)\gamma ^{\mu
}\right. \right.  \notag \\
&&\left. \left. \times S_{b}^{bb^{\prime }}(x)\right] \right] +\mathrm{Tr}%
\left[ \gamma _{\mu }\widetilde{S}_{c}^{b^{\prime }b}(-x)\gamma _{\nu
}S_{c}^{a^{\prime }a}(-x)\right]  \notag \\
&&\left. \times \left[ \mathrm{Tr}\left[ \gamma ^{\nu }\widetilde{S}%
_{b}^{aa^{\prime }}(x)\gamma ^{\mu }S_{b}^{bb^{\prime }}(x)\right] -\mathrm{%
Tr}\left[ \gamma ^{\nu }\widetilde{S}_{b}^{ba^{\prime }}(x)\gamma ^{\mu
}S_{b}^{ab^{\prime }}(x)\right] \right] \right\} .  \notag \\
&&  \label{eq:QCD1}
\end{eqnarray}%
Here%
\begin{equation}
\widetilde{S}_{Q}(x)=CS_{Q}^{T}(x)C,  \label{eq:Prop}
\end{equation}%
where $Q=b,\ c$, and $S_{b(c)}(x)$ are the $b$ and $c$-quark propagators.
They are given by the following formula
\begin{eqnarray}
&&S_{Q}^{ab}(x)=i\int \frac{d^{4}k}{(2\pi )^{4}}e^{-ikx}\Bigg \{\frac{\delta
_{ab}\left( {\slashed k}+m_{Q}\right) }{k^{2}-m_{Q}^{2}}  \notag \\
&&-\frac{g_{s}G_{ab}^{\alpha \beta }}{4}\frac{\sigma _{\alpha \beta }\left( {%
\slashed k}+m_{Q}\right) +\left( {\slashed k}+m_{Q}\right) \sigma _{\alpha
\beta }}{(k^{2}-m_{Q}^{2})^{2}}  \notag \\
&&+\frac{g_{s}^{2}G^{2}}{12}\delta _{ab}m_{Q}\frac{k^{2}+m_{Q}{\slashed k}}{%
(k^{2}-m_{Q}^{2})^{4}}+\cdots \Bigg \}.  \label{eq:QProp}
\end{eqnarray}%
In Eq.\ (\ref{eq:QProp}), we adopt the notations
\begin{equation}
G_{ab}^{\alpha \beta }\equiv G_{A}^{\alpha \beta }\lambda _{ab}^{A}/2,\ \
G^{2}=G_{\alpha \beta }^{A}G_{A}^{\alpha \beta },\
\end{equation}%
with $G_{A}^{\alpha \beta }$ being the gluon field-strength tensor. Above $%
\lambda ^{A}$ are the Gell-Mann matrices and index $A$ runs in the range $%
1\div 8$.

The $\Pi _{1}^{\mathrm{OPE}}(p)$ has also a trivial Lorentz structure which
is proportional to $\mathrm{I}$, and is characterized by the invariant
amplitude $\Pi _{1}^{\mathrm{OPE}}(p^{2})$. The SRs for the $m_{1}$ and $%
\Lambda _{1}$ are derived by equating the invariant amplitudes $\Pi _{1}^{%
\mathrm{OPE}}(p^{2})$ and $\Pi _{1}^{\mathrm{Phys}}(p^{2})$, applying the
Borel transformation to obtained equality and performing continuum
subtraction using the assumption on quark-hadron duality \cite%
{Shifman:1978bx,Shifman:1978by}. Then, it is not difficult to find that%
\begin{equation}
m_{1}^{2}=\frac{\Pi _{1}^{\prime }(M^{2},s_{0})}{\Pi _{1}(M^{2},s_{0})}
\label{eq:Mass}
\end{equation}%
and
\begin{equation}
\Lambda _{1}^{2}=e^{m_{1}^{2}/M^{2}}\Pi _{1}(M^{2},s_{0}),  \label{eq:Coupl}
\end{equation}%
where $\Pi _{1}(M^{2},s_{0})$ is the amplitude $\Pi _{1}^{\mathrm{OPE}%
}(p^{2})$ after the Borel transformation and continuum subtraction
operations. The quantities $M^{2}$ and $s_{0}$ are the Borel and continuum
subtraction parameters. In Eq.\ (\ref{eq:Mass}) a symbol $\Pi _{1}^{\prime
}(M^{2},s_{0})=d\Pi _{1}(M^{2},s_{0})/d(-1/M^{2})$ is used.

The amplitude $\Pi _{1}(M^{2},s_{0})$ has the form
\begin{equation}
\Pi _{1}(M^{2},s_{0})=\int_{4\mathcal{M}^{2}}^{s_{0}}ds\rho _{1}^{\mathrm{OPE%
}}(s)e^{-s/M^{2}},  \label{eq:InvAmp}
\end{equation}%
where $\mathcal{M=(}m_{b}+m_{c})$, and $\rho _{1}^{\mathrm{OPE}}(s)$ is a
two-point spectral density which is determined as an imaginary part of the
invariant amplitude $\Pi _{1}^{\mathrm{OPE}}(p^{2})$. The function $\rho
_{1}^{\mathrm{OPE}}(s)$ is a sum of a perturbative $\rho _{1}^{\mathrm{pert.}%
}(s)$ and dimension-$4$ nonperturbative $\rho _{1}^{\mathrm{Dim4}}(s)$
terms. Their explicit expressions are rather lengthy, therefore we do not
provide them here.

The SRs for the mass and current coupling of the tetraquark $X_{1}$ contain
the masses of the heavy quarks $m_{b}=4.18_{-0.02}^{+0.03}~\mathrm{GeV},$ $\
m_{c}=(1.27\pm 0.02)~\mathrm{GeV}$, as well as the gluon vacuum condensate $%
\langle \alpha _{s}G^{2}/\pi \rangle =(0.012\pm 0.004)~\mathrm{GeV}^{4}$.
The choice of the parameters $M^{2}$ and $s_{0}$ is another important
procedure in the SR computations, and should meet necessary restrictions
imposed on them by the method. Their choice has to guarantee dominance of
the pole contribution ($\mathrm{PC}$)
\begin{equation}
\mathrm{PC}=\frac{\Pi (M^{2},s_{0})}{\Pi (M^{2},\infty )},  \label{eq:PC}
\end{equation}%
to the correlation function $\Pi _{1}(M^{2},s_{0})$ and satisfy a constraint
$\mathrm{PC}\geq 0.5$. The highest $M^{2}$ which obeys this constraint fixes
the upper limit of $M^{2}$. The lower limit for the Borel parameter is
determined from convergence of $\mathrm{OPE}$. Because in our analysis there
is only a dimension-$4$ nonperturbative term, we choose $M^{2}$ in a such a
way that its contribution forms $\pm (5\div 15)\%$ of  $\Pi _{1}(M^{2},s_{0})
$. This ensures a prevalence of the perturbative term in $\Pi
_{1}(M^{2},s_{0})$. The stability of extracted quantities on the parameter $%
M^{2}$ is also among used restrictions.

\begin{widetext}

\begin{figure}[h!]
\begin{center}
\includegraphics[totalheight=6cm,width=8cm]{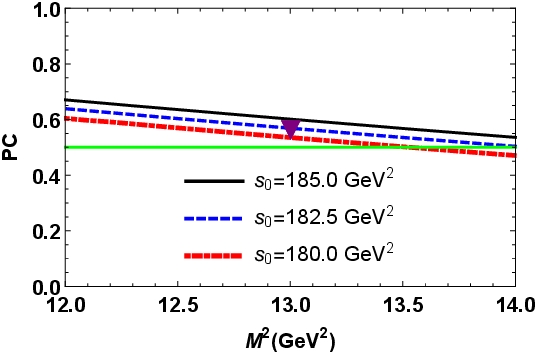}
\includegraphics[totalheight=6cm,width=8cm]{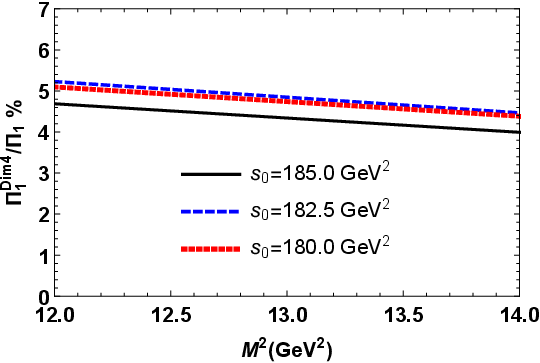}
\end{center}
\caption{Left: The pole contribution $\mathrm{PC}$ as a function of $M^{2}$.
The limit $\mathrm{PC}=0.5$ is shown by the green line. The red triangle shows the point, where the mass $m_1$
of the tetraquark $X_{\mathrm{1}}$ has been extracted from the sum rule
Right: The ratio $\Pi^{ \mathrm{Dim4}}_{1}(M^2,s_0)/\Pi_{1}(M^2,s_0)$
as a function
of the Borel parameter at fixed $s_0$.}
\label{fig:PCConv}
\end{figure}

\end{widetext}

Computations carried out by employing these constrains prove that the
regions
\begin{equation}
M^{2}\in \lbrack 12,14]~\mathrm{GeV}^{2},\ s_{0}\in \lbrack 180,185]~\mathrm{%
GeV}^{2},  \label{eq:Wind1}
\end{equation}%
comply with all aforementioned constraints. In fact, at $M^{2}=14~\mathrm{GeV%
}^{2}$ on the average in $s_{0}$ the pole contribution is equal to $\mathrm{%
PC}\approx 0.5$, whereas at $M^{2}=12~\mathrm{GeV}^{2}$ it amounts to $%
\approx 0.64$. The nonperturbative contribution is positive and at $M^{2}=12~%
\mathrm{GeV}^{2}$ constitutes only $4.5\%$ of the whole result. In Fig.\ \ref%
{fig:PCConv} (the left panel), we depict $\mathrm{PC}$ as a function of the
Borel parameter at the fixed $s_{0}$. As is seen, excluding only small
region for $s_{0}=180~\mathrm{GeV}^{2}$ at all values of $M^{2}$ and $s_{0}$
the pole contribution exceeds the $0.5$. The convergence of the $\mathrm{OPE}
$ and dominant nature of the perturbative contribution to $\Pi
_{1}(M^{2},s_{0})$ is seen in the right panel of the same Fig.\ \ref%
{fig:PCConv}.

The mass $m_{1}$ and current coupling $\Lambda _{1}$ of the tetraquark $X_{%
\mathrm{1}}$ are calculated as mean values of these parameters over the
regions Eq.\ (\ref{eq:Wind1}) and are equal to
\begin{eqnarray}
m_{1} &=&(12715\pm 80)~\mathrm{MeV},  \notag \\
\Lambda _{1} &=&(2.80\pm 0.29)~\mathrm{GeV}^{5}.  \label{eq:Result1}
\end{eqnarray}%
The predictions Eq.\ (\ref{eq:Result1}) correspond to SR results at the
point $M^{2}=13~\mathrm{GeV}^{2}$ and $s_{0}=182.5~\mathrm{GeV}^{2}$, where
the pole contribution is $\mathrm{PC}\approx 0.57$. This fact guarantees the
dominance of $\mathrm{PC}$ in the obtained results, and proves ground-level
nature of the tetraquark $X_{\mathrm{1}}$ in its class. Ambiguities in Eq.\ (%
\ref{eq:Result1}) are mainly generated by choices of the parameters $M^{2}$
and $s_{0}$. In the case of the mass $m_{1}$ they form only $\pm 0.6\%$ of
the obtained result which implies very high accuracy of performed analysis.
This fact can be explained by the sum rule for the mass Eq.\ (\ref{eq:Mass}%
), where it is defined as a ratio of two correlation functions. As a result,
variations in correlators are damped in $m_{1}$ which stabilizes numerical
output. In the case of $\Lambda _{1}$ ambiguities are equal to $\pm 10\%$
remaining within acceptable limits of SR analysis. The mass $m_{1}$ as a
function of $M^{2}$ and $s_{0}$ is plotted in Fig.\ \ref{fig:Mass1}.

\begin{widetext}

\begin{figure}[h!]
\begin{center}
\includegraphics[totalheight=6cm,width=8cm]{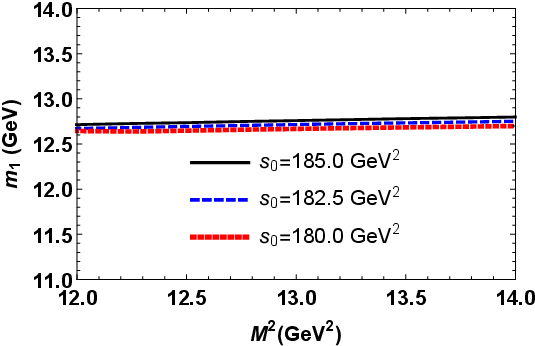}
\includegraphics[totalheight=6cm,width=8cm]{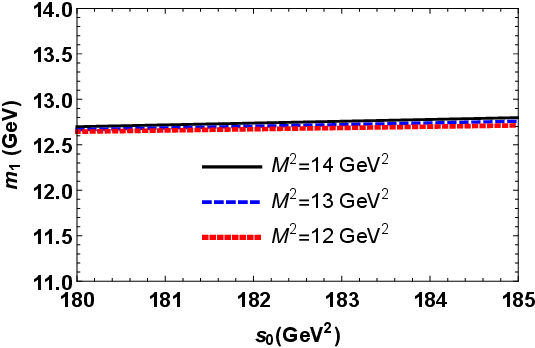}
\end{center}
\caption{Mass $m_1$ of the tetraquark $X_{\mathrm 1}$ as a function of the Borel parameter $M^{2}$ (left panel), and continuum threshold $s_0$ (right panel).}
\label{fig:Mass1}
\end{figure}

\begin{figure}[h!]
\begin{center}
\includegraphics[totalheight=6cm,width=8cm]{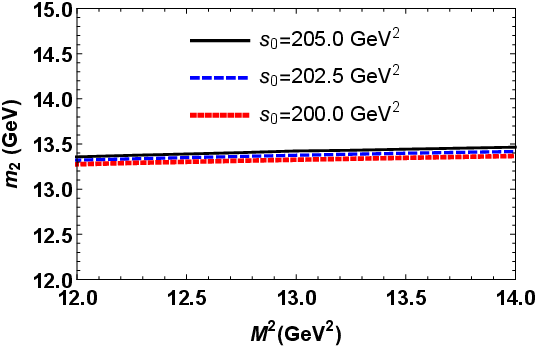}
\includegraphics[totalheight=6cm,width=8cm]{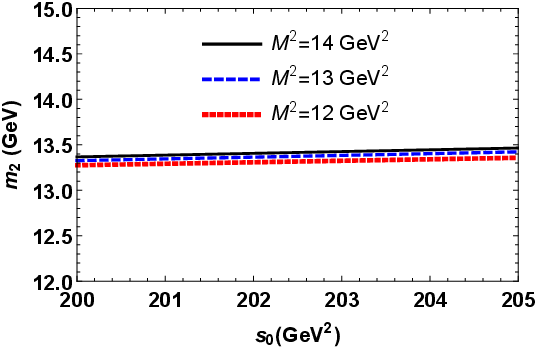}
\end{center}
\caption{Dependence of the mass $m_2$ on the Borel parameter $M^{2}$ (left panel), and continuum threshold parameter $s_0$ (right panel).}
\label{fig:Mass2}
\end{figure}

\end{widetext}

The mass $m_{2}$ and current coupling $\Lambda _{2}$ of the tetraquark $X_{%
\mathrm{2}}$ can be analyzed by the similar manner. The only difference here
is connected with the QCD side of the SRs. For the current $J_{2}(x)$ the
QCD side of the SR is given by the formula
\begin{eqnarray}
&&\Pi _{2}^{\mathrm{OPE}}(p)=i\int d^{4}xe^{ipx}\left\{ \mathrm{Tr}\left[
\widetilde{S}_{c}^{a^{\prime }b}(-x)S_{c}^{b^{\prime }a}(-x)\right] \right.
\notag \\
&&\times \left[ \mathrm{Tr}\left[ \widetilde{S}_{b}^{ba^{\prime
}}(x)S_{b}^{ab^{\prime }}(x)\right] +\mathrm{Tr}\left[ \widetilde{S}%
_{b}^{aa^{\prime }}(x)S_{b}^{bb^{\prime }}(x)\right] \right]  \notag \\
&&+\mathrm{Tr}\left[ \widetilde{S}_{c}^{b^{\prime }b}(-x)S_{c}^{a^{\prime
}a}(-x)\right] \left[ \mathrm{Tr}\left[ \widetilde{S}_{b}^{aa^{\prime
}}(x)S_{b}^{bb^{\prime }}(x)\right] \right.  \notag \\
&&\left. \left. +\mathrm{Tr}\left[ \widetilde{S}_{b}^{ba^{\prime
}}(x)S_{b}^{ab^{\prime }}(x)\right] \right] \right\} .  \label{eq:QCD2}
\end{eqnarray}%
The remaining manipulations required to evaluate the $m_{2}$ and $\Lambda
_{2}$ do not differ from ones explained above. Therefore, it is convenient
to provide final predictions for $m_{2}$ and $\Lambda _{2}$%
\begin{eqnarray}
m_{2} &=&(13370\pm 95)~\mathrm{MeV},  \notag \\
\Lambda _{2} &=&(1.19\pm 0.14)~\mathrm{GeV}^{5},  \label{eq:Result2}
\end{eqnarray}%
which are found using the working regions
\begin{equation}
M^{2}\in \lbrack 12,14]~\mathrm{GeV}^{2},\ s_{0}\in \lbrack 200,205]~\mathrm{%
GeV}^{2},  \label{eq:Wind2}
\end{equation}%
where all sum rule's constraints are satisfied. Indeed, the pole
contribution on average in $s_{0}$ is larger than $0.5$ for all values of
the Borel parameter and changes inside limits
\begin{equation}
0.66\geq \mathrm{PC}\geq 0.51
\end{equation}%
At $M^{2}=12~\mathrm{GeV}^{2}$ the nonperturbative contribution does not
exceed $4\%$ of the correlation function.\ Predictions extracted for the
mass $m_{2}$ of the tetraquark $X_{\mathrm{2}}$ are depicted in Fig.\ \ref%
{fig:Mass2}.


\section{Full width of the diquark-antidiquark state $X_{\mathrm{1}}$}

\label{sec:ScalarWidths1}


The masses $m_{1}$ and $m_{2}$ of the scalar tetraquarks $X_{\mathrm{1}}$
and $X_{\mathrm{2}}$ are important parameters to determine their possible
decay channels. It is evident that these particles can decay to pair of $%
\overline{c}b$ mesons with different masses and quantum numbers. Available
experimental information on these mesons is restricted by parameters of $%
B_{c}^{\pm }$ and $B_{c}^{\pm }(2S)$ \cite{PDG:2022}. Therefore, to explore
decays of $X_{\mathrm{1(2)}}$, we should use also results of theoretical
studies. The masses and decay constants of the numerous $\overline{c}b$
mesons were calculated in the framework of different methods. We are going
to employ the predictions $m_{B_{c}^{\ast }}=6338~\mathrm{MeV}$ and $%
f_{B_{c}^{\ast }}=471~\mathrm{MeV}$ for the mass and decay constant of the
vector particles $B_{c}^{\ast \pm }$ from Refs.\ \cite{Godfrey:2004ya} and
\cite{Eichten:2019gig}, respectively. It is easy \ to see that $m_{1}$
overshoots the thresholds $12549~\mathrm{MeV}$ and $12676~\mathrm{MeV}$ for
production of the pseudoscalar $2B_{c}^{-}$ and vector $2B_{c}^{\ast -}$%
mesons, respectively. At the same time, the mass of this particle is not
enough for the decay $X_{\mathrm{1}}\rightarrow B_{c}^{-}B_{c}^{-}(2S)$
where the mass limit is $13146~\mathrm{MeV}$.


\subsection{Decay $X_{\mathrm{1}}\rightarrow B_{c}^{-}B_{c}^{-}$}


We start from analysis of the process $X_{\mathrm{1}}\rightarrow
B_{c}^{-}B_{c}^{-}$. The width of this decay can be computed using the
strong coupling $g_{1}$ of particles at the vertex $X_{\mathrm{1}%
}B_{c}^{-}B_{c}^{-}$. For this purpose, there is a necessity to explore the
QCD three-point correlation function
\begin{eqnarray}
\Pi _{1}(p,p^{\prime }) &=&i^{2}\int d^{4}xd^{4}ye^{ip^{\prime
}y}e^{-ipx}\langle 0|\mathcal{T}\{J^{B_{c}}(y)  \notag \\
&&\times J^{B_{c}}(0)J_{1}^{\dagger }(x)\}|0\rangle ,  \label{eq:CF3}
\end{eqnarray}%
where
\begin{equation}
J^{B_{c}}(x)=\overline{c}_{i}(x)i\gamma _{5}b_{i}(x),  \label{eq:CR3}
\end{equation}%
is the interpolating current of the pseudoscalar meson $B_{c}^{-}$.

The correlator $\Pi _{1}(p,p^{\prime })$ allows one to find the SR for the
strong form factor $g_{1}(q^{2})$, which at the mass shell $%
q^{2}=m_{B_{c}}^{2}$ gives the coupling $g_{1}$. To derive SR for the form
factor $g_{1}(q^{2})$, we follow standard prescriptions of the method and
write down $\Pi _{1}(p,p^{\prime })$ using the parameters of the particles
involved into the decay process. The correlation function $\Pi _{1}^{\mathrm{%
Phys}}(p,p^{\prime })$ obtained by this manner constitutes the physical side
of the sum rule for the form factor $g_{1}(q^{2})$ and is given by the
formula
\begin{eqnarray}
&&\Pi _{1}^{\mathrm{Phys}}(p,p^{\prime })=\frac{\langle
0|J^{B_{c}}|B_{c}(p^{\prime })\rangle }{p^{\prime 2}-m_{B_{c}}^{2}}\frac{%
\langle 0|J^{B_{c}}|B_{c}(q)\rangle }{q^{2}-m_{B_{c}}^{2}}  \notag \\
&&\times \langle B_{c}(p^{\prime })B_{c}(q)|X_{\mathrm{1}}(p)\rangle \frac{%
\langle X_{\mathrm{1}}(p)|J_{1}^{\dagger }|0\rangle }{p^{2}-m_{1}^{2}}%
+\cdots ,  \label{eq:CF5}
\end{eqnarray}%
where $m_{B_{c}}=(6274.47\pm 0.27)~\mathrm{MeV}$ is the mass of $B_{c}^{-}$
meson. The correlator $\Pi _{1}^{\mathrm{Phys}}(p,p^{\prime })$ is found
after isolating a contribution of the ground-state particles, whereas
effects due to higher states and continuum are denoted by the ellipses. To
recast $\Pi _{1}^{\mathrm{Phys}}(p,p^{\prime })$ to easy-to-use form, we
employ the matrix element
\begin{equation}
\langle 0|J^{B_{c}}|B_{c}\rangle =\frac{f_{B_{c}}m_{B_{c}}^{2}}{m_{b}+m_{c}},
\label{eq:ME2}
\end{equation}%
with $f_{B_{c}}=(476\pm 27)~\mathrm{MeV}$ being the decay constant of the
meson $B_{c}^{\pm }$ \cite{Veliev:2010vd}. We also model the vertex $X_{%
\mathrm{1}}B_{c}B_{c}$ using the expression%
\begin{equation}
\langle B_{c}(p^{\prime })B_{c}(q)|X_{\mathrm{1}}(p)\rangle
=g_{1}(q^{2})p\cdot p^{\prime }.  \label{eq:ME3}
\end{equation}%
Then, the correlation function (\ref{eq:CF5}) can be presented in the
following form
\begin{eqnarray}
&&\Pi _{1}^{\mathrm{Phys}}(p,p^{\prime })=g_{1}(q^{2})\frac{\Lambda
_{1}f_{B_{c}}^{2}m_{B_{c}}^{4}}{(m_{b}+m_{c})^{2}\left(
p^{2}-m_{1}^{2}\right) }  \notag \\
&&\times \frac{m_{1}^{2}+m_{B_{c}}^{2}-q^{2}}{2\left( p^{\prime
2}-m_{B_{c}}^{2}\right) (q^{2}-m_{B_{c}}^{2})}+\cdots .  \label{eq:CF6}
\end{eqnarray}%
Because $\Pi _{1}^{\mathrm{Phys}}(p,p^{\prime })$ has a simple Lorentz
structure proportional to $\mathrm{I}$, we denote by $\Pi _{1}^{\mathrm{Phys}%
}(p^{2},p^{\prime 2},q^{2})$ the right-hand side of Eq.\ (\ref{eq:CF6}) as
the invariant amplitude and use it to derive SR for $g_{1}(q^{2})$.

The QCD side of SR is equal to
\begin{eqnarray}
&&\Pi _{1}^{\mathrm{OPE}}(p,p^{\prime })=2i^{2}\int
d^{4}xd^{4}ye^{ip^{\prime }y}e^{-ipx}\left\{ \mathrm{Tr}\left[ \gamma
_{5}S_{b}^{ib}(y-x)\right. \right.   \notag \\
&&\left. \times \gamma _{\mu }\widetilde{S}_{b}^{ja}(-x)\gamma _{5}%
\widetilde{S}_{c}^{bj}(x)\gamma ^{\mu }S_{c}^{ai}(x-y)\right]   \notag \\
&&\left. -\mathrm{Tr}\left[ \gamma _{5}S_{b}^{ib}(y-x)\gamma _{\mu }%
\widetilde{S}_{b}^{ja}(-x){}\gamma _{5}\widetilde{S}_{c}^{aj}(x)\gamma ^{\mu
}S_{c}^{bi}(x-y)\right] \right\} .  \notag \\
&&  \label{eq:QCDside2}
\end{eqnarray}%
The $\Pi _{1}^{\mathrm{OPE}}(p,p^{\prime })$ has also a trivial Lorentz
structure, and $\Pi _{1}^{\mathrm{OPE}}(p^{2},p^{\prime 2},q^{2})$ stands
for the corresponding amplitude. Then, the SR for the form factor $%
g_{1}(q^{2})$ reads%
\begin{eqnarray}
&&g_{1}(q^{2})=\frac{2(m_{b}+m_{c})^{2}}{\Lambda
_{1}f_{B_{c}}^{2}m_{B_{c}}^{4}}\frac{q^{2}-m_{B_{c}}^{2}}{%
m^{2}+m_{B_{c}}^{2}-q^{2}}  \notag \\
&&\times e^{m_{1}^{2}/M_{1}^{2}}e^{m_{B_{c}}^{2}/M_{2}^{2}}\Pi _{1}(\mathbf{M%
}^{2},\mathbf{s}_{0},q^{2}).  \label{eq:SRCoup2}
\end{eqnarray}%
Here,
\begin{eqnarray}
&&\Pi _{1}(\mathbf{M}^{2},\mathbf{s}_{0},q^{2})=\int_{4\mathcal{M}%
^{2}}^{s_{0}}ds\int_{\mathcal{M}^{2}}^{s_{0}^{\prime }}ds^{\prime }\rho
_{1}(s,s^{\prime },q^{2})  \notag \\
&&\times e^{-s/M_{1}^{2}}e^{-s^{\prime }/M_{2}^{2}}.
\end{eqnarray}%
is the function $\Pi _{1}^{\mathrm{OPE}}(p^{2},p^{\prime 2},q^{2})$
undergone Borel transformations and continuum subtractions. It is written in
term of the spectral density $\rho (s,s^{\prime },q^{2})$. In Eq.\ (\ref%
{eq:SRCoup2}) $\mathbf{M}^{2}=(M_{1}^{2},M_{2}^{2})$ and $\mathbf{s}%
_{0}=(s_{0},s_{0}^{\prime })$ are the Borel and continuum threshold
parameters, respectively.

To carry out numerical computations, we use for $M_{1}^{2}$ and $s_{0}$,
associated with the $X_{\mathrm{1}}$ channel, the regions from Eq.\ (\ref%
{eq:Wind1}). The parameters $(M_{2}^{2},\ s_{0}^{\prime })$ for the $%
B_{c}^{-}$ meson channel are changed within the limits%
\begin{equation}
M_{2}^{2}\in \lbrack 6.5,7.5]~\mathrm{GeV}^{2},\ s_{0}^{\prime }\in \lbrack
45,47]~\mathrm{GeV}^{2}.  \label{eq:Wind3}
\end{equation}

The SR approach generates reliable predictions for the form factor $%
g_{1}(q^{2})$ in the Euclidean region $q^{2}<0$. But the strong coupling $%
g_{1}$should be determined at the mass shell $q^{2}=m_{B_{c}}^{2}$. To evade
this obstacle, it is appropriate to introduce a new variable $Q^{2}=-q^{2}$
and use $g_{1}(Q^{2})$ for the obtained function. Afterwards, we employ a
function $\mathcal{F}_{1}(Q^{2})$ that at momenta $Q^{2}>0$ generates values
coinciding with the SR predictions, but can be extrapolated to the region of
$Q^{2}<0$. For these purposes, we employ the functions $\mathcal{F}%
_{i}(Q^{2},m_{1}^{2})$
\begin{equation}
\mathcal{F}_{i}(Q^{2},m_{1}^{2})=\mathcal{F}_{i}^{0}\mathrm{\exp }\left[
c_{i}^{1}\frac{Q^{2}}{m_{1}^{2}}+c_{i}^{2}\left( \frac{Q^{2}}{m_{1}^{2}}%
\right) ^{2}\right]
\end{equation}%
with parameters $\mathcal{F}_{i}^{0}$, $c_{i}^{1}$, and $c_{i}^{2}$.

\begin{figure}[h]
\includegraphics[width=8.5cm]{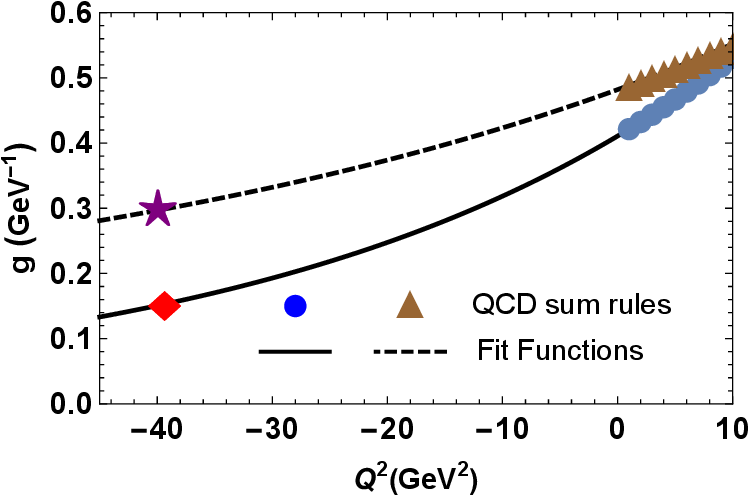}
\caption{QCD data and fit functions for the form factors $g_{1}(Q^{2})$
(solid line) and $g_{2}(Q^{2})$ (dashed line). The diamond on the solid line
and star on the dashed curve fix the points $Q^{2}=-m_{B_c}^{2}$ and $%
Q^{2}=-m_{B_{c}^{\ast}}^{2}$ where $g_{1}$ and $g_{2}$ have been extracted. }
\label{fig:Fit}
\end{figure}

In our present analysis, SR calculations comprise $Q^{2}$ from the interval $%
Q^{2}=1\div 10~\mathrm{GeV}^{2}$. Results of relevant computations are
plotted in Fig.\ \ref{fig:Fit}. It is easy to fix the parameters $\mathcal{F}%
_{1}^{0}=0.41~\mathrm{GeV}^{-1}$, $c_{1}^{1}=4.13$, and $c_{1}^{2}=0.31$ of
the function $\mathcal{F}_{1}(Q^{2},m_{1}^{2})$ which leads to reasonable
agreement with the SR data: This function is shown in Fig.\ \ref{fig:Fit} as
well.

Then, for the strong coupling $g_{1}$, we get
\begin{equation}
g_{1}\equiv \mathcal{F}_{1}(-m_{B_{c}}^{2},m_{1}^{2})=(1.7\pm 0.3)\times
10^{-1}\ \mathrm{GeV}^{-1}.
\end{equation}%
The width of the process $X_{\mathrm{1}}\rightarrow B_{c}^{-}B_{c}^{-}$ is
given by the formula%
\begin{equation}
\Gamma \left[ X_{\mathrm{1}}\rightarrow B_{c}^{-}B_{c}^{-}\right] =g_{1}^{2}%
\frac{m_{B_{c}}^{2}\lambda _{1}}{8\pi }\left( 1+\frac{\lambda _{1}^{2}}{%
m_{B_{c}}^{2}}\right) ,  \label{eq:PDw2}
\end{equation}%
where $\lambda _{1}=\lambda (m_{1},m_{B_{c}},m_{B_{c}})$, and
\begin{equation}
\lambda (x,y,z)=\frac{\sqrt{%
x^{4}+y^{4}+z^{4}-2(x^{2}y^{2}+x^{2}z^{2}+y^{2}z^{2})}}{2x}.
\end{equation}%
As a result, we obtain
\begin{equation}
\Gamma \left[ X_{\mathrm{1}}\rightarrow B_{c}^{-}B_{c}^{-}\right] =(38.3\pm
9.8)~\mathrm{MeV}.  \label{eq:DW1}
\end{equation}

\subsection{Process $X_{\mathrm{1}}\rightarrow B_{c}^{\ast -}B_{c}^{\ast -}$}


The decay $X_{\mathrm{1}}\rightarrow B_{c}^{\ast -}B_{c}^{\ast -}$ is $S$%
-wave process, partial width of which is determined by the strong coupling $%
g_{2}$ at the vertex $X_{\mathrm{1}}B_{c}^{\ast -}B_{c}^{\ast -}$. In the
context of the QCD SR method the form factor $g_{2}(q^{2})$ is calculated
using the three-point correlation function%
\begin{eqnarray}
&&\Pi _{\mu \nu }(p,p^{\prime })=i^{2}\int d^{4}xd^{4}ye^{ip^{\prime
}y}e^{-ipx}\langle 0|\mathcal{T}\{J_{\mu }^{B_{c}^{\ast }}(y)  \notag \\
&&\times J_{\nu }^{B_{c}^{\ast }}(0)J_{1}^{\dagger }(x)\}|0\rangle ,
\label{eq:CF7}
\end{eqnarray}%
with $J_{\mu }^{B_{c}^{\ast }}(x)$ being the interpolating current for the
vector meson $B_{c}^{\ast -}$
\begin{equation}
J_{\mu }^{B_{c}^{\ast }}(x)=\overline{c}_{i}(x)\gamma _{\mu }b_{i}(x).
\label{eq:CR4}
\end{equation}%
Here, $p$, $p^{\prime }$, and $q=p-p^{\prime }$ are $4$-momenta of the
tetraquark $X_{\mathrm{1}}(p)$, and $B_{c}^{\ast -}(p^{\prime })$, $%
B_{c}^{\ast -}(q)$ mesons, respectively.

To compute the phenomenological side of the SR, we need the following matrix
elements
\begin{equation}
\langle 0|J_{\mu }^{B_{c}^{\ast }}|B_{c}^{\ast }(p)\rangle =f_{B_{c}^{\ast
}}m_{B_{c}^{\ast }}\varepsilon _{\mu }(p),  \label{eq:ME2A}
\end{equation}%
and
\begin{eqnarray}
&&\langle B_{c}^{\ast }(p^{\prime })B_{c}^{\ast }(q)|X_{\mathrm{1}%
}(p)\rangle =g_{2}(q^{2})\left[ q\cdot p^{\prime }\varepsilon ^{\ast
}(p^{\prime })\cdot \varepsilon ^{\ast }(q)\right.  \notag \\
&&\left. -q\cdot \varepsilon ^{\ast }(p^{\prime })p^{\prime }\cdot
\varepsilon ^{\ast }(q)\right] ,  \label{eq:ME3A}
\end{eqnarray}%
where $\varepsilon (p^{\prime })$ and $\varepsilon (q)$ are the polarization
vectors of $B_{c}^{\ast -}(p^{\prime })$, and $B_{c}^{\ast -}(q)$.

After some computations, we get for the physical side of the sum rule
\begin{eqnarray}
&&\Pi _{\mu \nu }^{\mathrm{Phys}}(p,p^{\prime })=g_{2}(q^{2})\frac{\Lambda
_{1}f_{B_{c}^{\ast }}^{2}m_{B_{c}^{\ast }}^{2}}{\left(
p^{2}-m_{1}^{2}\right) \left( p^{\prime 2}-m_{B_{c}^{\ast }}^{2}\right)
(q^{2}-m_{B_{c}^{\ast }}^{2})}  \notag \\
&&\times \left[ \frac{1}{2}\left( m_{1}^{2}-m_{B_{c}^{\ast
}}^{2}-q^{2}\right) g_{\mu \nu }-q_{\mu }p_{\nu }^{\prime }\right] +\cdots .
\label{eq:CR2A}
\end{eqnarray}

The correlator $\Pi _{\mu \nu }^{\mathrm{Phys}}(p,p^{\prime })$ contains two
Lorentz structures that can be used to obtain the SR for $g_{2}(q^{2})$. We
select to work with the contribution $\sim g_{\mu \nu }$ and present the
corresponding invariant amplitude as $\widehat{\Pi }_{1}^{\mathrm{Phys}%
}(p^{2},p^{\prime 2},q^{2})$. The Borel transformations of the amplitude $%
\widehat{\Pi }_{1}^{\mathrm{Phys}}(p^{2},p^{\prime 2},q^{2})$ over $-p^{2}$
and $-p^{\prime 2}$ yield
\begin{eqnarray}
&&\mathcal{B}\widehat{\Pi }_{1}^{\mathrm{Phys}}(p^{2},p^{\prime
2},q^{2})=g_{2}(q^{2})\Lambda _{1}f_{B_{c}^{\ast }}^{2}m_{B_{c}^{\ast }}^{2}
\notag \\
&&\times \frac{m_{1}^{2}-m_{B_{c}^{\ast }}^{2}-q^{2}}{2(q^{2}-m_{B_{c}^{\ast
}}^{2})}e^{-m_{1}^{2}/M_{1}^{2}}e^{-m_{B_{c}^{\ast }}^{2}/M_{2}^{2}}+\cdots .
\label{eq:CorrF5a}
\end{eqnarray}

The correlation function $\Pi _{\mu \nu }(p,p^{\prime })$ expressed in terms
of $c$-quark propagators reads
\begin{eqnarray}
&&\Pi _{\mu \nu }^{\mathrm{OPE}}(p,p^{\prime })=2i^{2}\int
d^{4}xd^{4}ye^{ip^{\prime }y}e^{-ipx}\left\{ \mathrm{Tr}\left[ \gamma _{\mu
}S_{b}^{ib}(y-x)\right. \right.   \notag \\
&&\left. \times \gamma _{\theta }\widetilde{S}_{b}^{ja}(-x){}\gamma _{\nu }%
\widetilde{S}_{c}^{bj}(x)\gamma ^{\theta }S_{c}^{ai}(x-y)\right]   \notag \\
&&\left. -\mathrm{Tr}\left[ \gamma _{\mu }S_{b}^{ib}(y-x)\gamma _{\theta }%
\widetilde{S}_{b}^{ja}(-x){}\gamma _{\nu }\widetilde{S}_{c}^{aj}(x)\gamma
^{\theta }S_{c}^{bi}(x-y)\right] \right\} .  \notag \\
&&  \label{eq:QCDside}
\end{eqnarray}%
The invariant amplitude $\widehat{\Pi }_{1}^{\mathrm{OPE}}(p^{2},p^{\prime
2},q^{2})$ corresponding to the component $\sim g_{\mu \nu }$ in Eq.\ (\ref%
{eq:QCDside}) forms the QCD side of the SR.

Having equated amplitudes $\widehat{\Pi }_{1}^{\mathrm{OPE}}(p^{2},p^{\prime
2},q^{2})$ and $\widehat{\Pi }_{1}^{\mathrm{Phys}}(p^{2},p^{\prime 2},q^{2})$
and performed the doubly Borel transforms and continuum subtractions, one
can find the sum rule for the form factor $g_{2}(q^{2})$%
\begin{eqnarray}
&&g_{2}(q^{2})=\frac{2}{\Lambda _{1}f_{B_{c}^{\ast }}^{2}m_{B_{c}^{\ast
}}^{2}}\frac{q^{2}-m_{B_{c}^{\ast }}^{2}}{m_{1}^{2}-m_{B_{c}^{\ast
}}^{2}-q^{2}}  \notag \\
&&\times e^{m_{1}^{2}/M_{1}^{2}}e^{m_{B_{c}^{\ast }}^{2}/M_{2}^{2}}\widehat{%
\Pi }_{1}(\mathbf{M}^{2},\mathbf{s}_{0},q^{2}).  \label{eq:SRCoup}
\end{eqnarray}%
The remaining manipulations do not differ considerably from ones described
above. Therefore, let us write down the region for parameters $M_{2}^{2}$,
and $s_{0}^{\prime }$ in the channel of $B_{c}^{\ast -}$ meson
\begin{equation}
M_{2}^{2}\in \lbrack 6.5,7.5]~\mathrm{GeV}^{2},\ s_{0}^{\prime }\in \lbrack
49,51]~\mathrm{GeV}^{2}.
\end{equation}%
The extrapolating function $\mathcal{F}_{2}(Q^{2},m_{1}^{2})$ has the
parameters: $\mathcal{F}_{2}^{0}=0.48~\mathrm{GeV}^{-1}$, $c_{2}^{1}=2.14$,
and $c_{2}^{2}=0.74$. Then, the strong coupling $g_{2}$ is equal to
\begin{equation}
g_{2}\equiv \mathcal{F}_{2}(-m_{B_{c}^{\ast }}^{2},m_{1}^{2})=(3.0\pm
0.4)\times 10^{-1}\ \mathrm{GeV}^{-1}.
\end{equation}%
The width of the decay $X_{\mathrm{1}}\rightarrow B_{c}^{\ast -}B_{c}^{\ast
-}$ can be evaluated by means of the expression%
\begin{equation}
\Gamma \left[ X_{\mathrm{1}}\rightarrow B_{c}^{\ast -}B_{c}^{\ast -}\right]
=g_{2}^{2}\frac{\lambda _{2}}{8\pi }\left( \frac{m_{B_{c}^{\ast }}^{4}}{%
m_{1}^{2}}+\frac{2\lambda _{2}^{2}}{3}\right) ,  \label{eq:PartDW}
\end{equation}%
where $\lambda _{2}=\lambda (m_{1},m_{B_{c}^{\ast }},m_{B_{c}^{\ast }})$.

Having used Eq.\ (\ref{eq:PartDW}) and the strong coupling $g_{2}$, we find
\begin{equation}
\Gamma \left[ X_{\mathrm{1}}\rightarrow B_{c}^{\ast -}B_{c}^{\ast -}\right]
=(24.3\pm 6.2)~\mathrm{MeV}.  \label{eq:DW2}
\end{equation}%
Two processes considered in this section form the full width of the
tetraquark $X_{\mathrm{1}}$ which amounts to
\begin{equation}
\Gamma _{\mathrm{1}}=(63\pm 12)~\mathrm{MeV}.
\end{equation}


\section{Full width of the tetraquark $X_{\mathrm{2}}$}

\label{sec:ScalarWidths2}


The prediction for the mass $m_{2}=13370~\mathrm{MeV}$ of the state $X_{%
\mathrm{2}}$ demonstrates that it is considerably heavier than $X_{\mathrm{1}%
}$. As a result, $X_{\mathrm{2}}$ can decay to $2B_{c}^{-}$, and $%
2B_{c}^{\ast -}$ mesons, but its mass is also enough for the process $X_{%
\mathrm{2}}\rightarrow B_{c}^{-}B_{c}^{-}(2S)$ with the mass threshold of $%
13146~\mathrm{MeV}$.


\subsection{$X_{\mathrm{2}}\rightarrow B_{c}^{-}B_{c}^{-}$ and $X_{\mathrm{2}%
}\rightarrow B_{c}^{-}B_{c}^{-}(2S)$}


Because the mesons $B_{c}^{-}$ and $B_{c}^{-}(2S)$ are described by the same
interpolating current these two decays should be treated in a correlated
form. The correlation function required to extract the strong form factors $%
G_{1}(q^{2})$ and $G_{2}(q^{2})$ at the vertices $X_{\mathrm{2}%
}B_{c}^{-}B_{c}^{-}$ and $X_{\mathrm{2}}B_{c}^{-}B_{c}^{-}(2S)$ is given by
the expression
\begin{eqnarray}
\Pi _{2}(p,p^{\prime }) &=&i^{2}\int d^{4}xd^{4}ye^{ip^{\prime
}y}e^{-ipx}\langle 0|\mathcal{T}\{J^{B_{c}}(y)  \notag \\
&&\times J^{B_{c}}(0)J_{2}^{\dagger }(x)\}|0\rangle .  \label{eq:CF8}
\end{eqnarray}%
The phenomenological side of SRs for these form factors is determined by the
formula
\begin{eqnarray}
&&\Pi _{2}^{\mathrm{Phys}}(p,p^{\prime })=G_{1}(q^{2})\frac{\Lambda
_{2}f_{B_{c}}^{2}m_{B_{c}}^{4}}{(m_{b}+m_{c})^{2}\left(
p^{2}-m_{2}^{2}\right) }  \notag \\
&&\times \frac{m_{2}^{2}+m_{B_{c}}^{2}-q^{2}}{2\left( p^{\prime
2}-m_{B_{c}}^{2}\right) (q^{2}-m_{B_{c}}^{2})}  \notag \\
&&+G_{2}(q^{2})\frac{\Lambda _{2}\widetilde{f}_{B_{c}}\widetilde{m}%
_{B_{c}}^{2}f_{B_{c}}m_{B_{c}}^{2}}{(m_{b}+m_{c})^{2}\left(
p^{2}-m_{2}^{2}\right) }  \notag \\
&&\times \frac{m_{2}^{2}+\widetilde{m}_{B_{c}}^{2}-q^{2}}{2\left( p^{\prime
2}-\widetilde{m}_{B_{c}}^{2}\right) (q^{2}-m_{B_{c}}^{2})}+\cdots ,
\label{eq:CorrF5}
\end{eqnarray}%
where $\widetilde{m}_{B_{c}}=(6871.2\pm 1.0)~\mathrm{MeV}$ and $\widetilde{f}%
_{B_{c}}=(420\pm 20)~\mathrm{MeV}$ are the mass and decay constant of the
radially excited meson $B_{c}^{-}(2S)$ from Refs.\ \cite{PDG:2022} and \cite%
{Aliev:2019wcm}, respectively.

The second component of the sum rules, i.e., the correlation function $\Pi
_{2}(p,p^{\prime })$ in terms of quark propagators takes the following form:%
\begin{eqnarray}
&&\Pi _{2}^{\mathrm{OPE}}(p,p^{\prime })=2i^{2}\int
d^{4}xd^{4}ye^{ip^{\prime }y}e^{-ipx}\left\{ \mathrm{Tr}\left[ \gamma
_{5}S_{b}^{ib}(y-x)\right. \right.  \notag \\
&&\left. \times \widetilde{S}_{b}^{ja}(-x)\gamma _{5}\widetilde{S}%
_{c}^{aj}(x)S_{c}^{bi}(x-y)\right]  \notag \\
&&\left. +\mathrm{Tr}\left[ \gamma _{5}S_{b}^{ib}(y-x)\widetilde{S}%
_{b}^{ja}(-x){}\gamma _{5}\widetilde{S}_{c}^{bj}(x)S_{c}^{ai}(x-y)\right]
\right\} .  \notag \\
&&  \label{eq:CF8A}
\end{eqnarray}%
Functions $\Pi _{2}^{\mathrm{Phys}}(p,p^{\prime })$ and $\Pi _{2}^{\mathrm{%
OPE}}(p,p^{\prime })$ have trivial Lorentz structures. Having denoted
corresponding invariant amplitudes by $\Pi _{2}^{\mathrm{Phys}%
}(p^{2},p^{\prime 2},q^{2})$ and $\Pi _{2}^{\mathrm{OPE}}(p^{2},p^{\prime
2},q^{2})$, equated them to each other and carried out usual operations, it
is possible to derive the required SR equality for the form factors $%
G_{1}(q^{2})$ and $G_{2}(q^{2})$.

An expression obtained by this way depends on two unknown functions $%
G_{1}(q^{2})$ and $G_{2}(q^{2})$. We solve this problem gradually: At the
first step, we consider the form factor $G_{1}(q^{2})$, and use it at the
next stage to determine $G_{2}(q^{2})$. These two stages differ from each
another by a choice of the regions for the parameters $(M_{2}^{2},s_{0}^{%
\prime })$. At the first phase of analysis, we restrict $s_{0}^{\prime }$ by
the mass of the orbitally excited meson $B_{c}^{-}(2S)$ choosing $%
s_{0}^{\prime }<\widetilde{m}_{B_{c}}^{2}$. This enable us to include
contribution of the vertex $X_{\mathrm{2}}B_{c}^{-}B_{c}^{-}(2S)$ into the
"continuum" and consider only the first term in Eq.\ (\ref{eq:CorrF5}). The
relevant parameters $(M_{2}^{2},s_{0}^{\prime })$ are given in Eq.\ (\ref%
{eq:Wind3}).

At the next step, we fix
\begin{equation}
M_{2}^{2}\in \lbrack 6.5,7.5]~\mathrm{GeV}^{2}\text{, }s_{0}^{\ast ^{\prime
}}\in \lbrack 48,50]~\mathrm{GeV}^{2},  \label{eq:Wind3A}
\end{equation}%
and use $G_{1}(q^{2})$ as input information to extract the form factor $%
G_{2}(q^{2})$, where $s_{0}^{\ast ^{\prime }}$ is less than $%
m^{2}[B_{c}(3S)]=(7.272)^{2}~\mathrm{GeV}^{2}$ \cite{Godfrey:2004ya}. In
both steps for $(M_{1}^{2},s_{0})$ we employ the parameters from Eq.\ (\ref%
{eq:Wind2}).

Analysis performed in the framework of this approach leads to the results
\begin{equation}
G_{1}\equiv \mathcal{G}_{1}(-m_{B_{c}}^{2},m_{2}^{2})=(1.1\pm 0.2)\times
10^{-1}\ \mathrm{GeV}^{-1},
\end{equation}%
and
\begin{equation}
G_{2}\equiv \mathcal{G}_{2}(-m_{B_{c}}^{2},m_{2}^{2})=(0.7\pm 0.1)\times
10^{-1}\ \mathrm{GeV}^{-1},
\end{equation}%
where $\mathcal{G}_{i}(Q^{2}$,$m_{2}^{2})$ have the same analytic form as $%
\mathcal{F}_{i}(Q^{2}$,$m_{1}^{2})$ but with substitution $%
m_{1}^{2}\rightarrow m_{2}^{2}$. Let us note that though the couplings $%
G_{1} $ and $G_{2}$ are determined at the mass shell of the $B_{c}^{-}$
meson, due to fitting parameters $\mathcal{G}_{1}$ and $\mathcal{G}_{2}$ are
different functions.

The width of the decays $X_{\mathrm{2}}\rightarrow B_{c}^{-}B_{c}^{-}$ and $%
X_{\mathrm{2}}\rightarrow B_{c}^{-}B_{c}^{-}(2S)$ can be computed using Eq.\
(\ref{eq:PDw2}) after relevant replacements

\begin{eqnarray}
&&\Gamma \left[ X_{\mathrm{2}}\rightarrow B_{c}^{-}B_{c}^{-}\right]
=(45.5\pm 11.7)~\mathrm{MeV},  \notag \\
&&\Gamma \left[ X_{\mathrm{2}}\rightarrow B_{c}^{-}B_{c}^{-}(2S)\right]
=(11.9\pm 2.8)~\mathrm{MeV}.
\end{eqnarray}


\subsection{$X_{\mathrm{2}}\rightarrow B_{c}^{\ast -}B_{c}^{\ast -}$}


We are going to analyze the decay $X_{\mathrm{2}}\rightarrow B_{c}^{\ast
-}B_{c}^{\ast -}$ by means of the correlation function

\begin{eqnarray}
\widetilde{\Pi }_{\mu \nu }(p,p^{\prime }) &=&i^{2}\int
d^{4}xd^{4}ye^{ip^{\prime }y}e^{-ipx}\langle 0|\mathcal{T}\{J_{\mu
}^{B_{c}^{\ast }}(y)  \notag \\
&&\times J_{\nu }^{B_{c}^{\ast }}(0)J_{2}^{\dagger }(x)\}|0\rangle .
\label{eq:CF9}
\end{eqnarray}%
Using the parameters $m_{2}$, $\Lambda _{2}$ and $f_{B_{c}^{\ast }}$, $%
m_{B_{c}^{\ast }}$, one can recast the correlator $\widetilde{\Pi }_{\mu \nu
}(p,p^{\prime })$ into the form
\begin{eqnarray}
&&\widetilde{\Pi }_{\mu \nu }^{\mathrm{Phys}}(p,p^{\prime })=\frac{%
G_{3}(q^{2})\Lambda _{2}f_{B_{c}^{\ast }}^{2}m_{B_{c}^{\ast }}^{2}}{\left(
p^{2}-m_{2}^{2}\right) \left( p^{\prime 2}-m_{B_{c}^{\ast }}^{2}\right)
(q^{2}-m_{B_{c}^{\ast }}^{2})}  \notag \\
&&\times \left[ \frac{1}{2}\left( m_{2}^{2}-m_{B_{c}^{\ast
}}^{2}-q^{2}\right) g_{\mu \nu }-q_{\mu }p_{\nu }^{\prime }\right] +\cdots ,
\end{eqnarray}%
where $G_{3}(q^{2})$ is the strong form factor of interest. Evidently, $%
\widetilde{\Pi }_{\mu \nu }^{\mathrm{Phys}}(p,p^{\prime })$ can be used to
extract the invariant amplitude necessary for our purposes. We work with the
Lorentz structure $g_{\mu \nu }$, therefore relevant part of $\widetilde{\Pi
}_{\mu \nu }^{\mathrm{Phys}}(p,p^{\prime })$ constitutes the physical side
of SR. The sum rule's QCD side is determined by the function
\begin{eqnarray}
&&\widetilde{\Pi }_{\mu \nu }^{\mathrm{OPE}}(p,p^{\prime })=2i^{2}\int
d^{4}xd^{4}ye^{ip^{\prime }y}e^{-ipx}\left\{ \mathrm{Tr}\left[ \gamma _{\mu
}S_{b}^{ib}(y-x)\right. \right.  \notag \\
&&\left. \times \widetilde{S}_{b}^{ja}(-x){}\gamma _{\nu }\widetilde{S}%
_{c}^{bj}(x)S_{c}^{ai}(x-y)\right]  \notag \\
&&\left. +\mathrm{Tr}\left[ \gamma _{\mu }S_{b}^{ib}(y-x)\widetilde{S}%
_{b}^{ja}(-x){}\gamma _{\nu }\widetilde{S}_{c}^{aj}(x)S_{c}^{bi}(x-y)\right]
\right\} .
\end{eqnarray}

Having omitted further details, we write down the final results for the
parameters of this decay:%
\begin{equation}
G_{3}\equiv \mathcal{G}_{3}(-m_{B_{c}^{\ast }}^{2},m_{2}^{2})=(1.4\pm
0.3)\times 10^{-1}\ \mathrm{GeV}^{-1},
\end{equation}%
and
\begin{equation}
\Gamma \left[ X_{\mathrm{2}}\rightarrow B_{c}^{\ast -}B_{c}^{\ast -}\right]
=(21.3\pm 6.5)~\mathrm{MeV}.
\end{equation}%
Then, the full width of the tetraquark $X_{\mathrm{2}}$ is
\begin{equation}
\Gamma _{2}=(79\pm 14)~\mathrm{MeV}.
\end{equation}


\section{Analysis and conclusions}

\label{sec:Conc}


In present work, we have investigated the doubly charged scalar
diquark-antidiquark states $X_{\mathrm{1}}$ and $X_{\mathrm{2}}$ with the
axial-vector and pseudoscalar ingredients, respectively. The tetraquark $X_{%
\mathrm{1}}$ is the $\overline{\mathbf{3}_{c}}\otimes \mathbf{3}_{c}$ color
state, and is expected to occupy one of a lowest level in the spectroscopy
of similar exotic mesons. The second diquark-antidiquark state $X_{\mathrm{2}%
}$ studied here has the same quantum numbers, but different inner structure:
it is built of color-sextet diquarks $\mathbf{6}_{c}\otimes \overline{%
\mathbf{6}}_{c}$.

The mass $m_{1}=(12715\pm 80)~\mathrm{MeV}$ of $X_{\mathrm{1}}$ is
considerably smaller than the mass of the structure $X_{\mathrm{2}}$. But $%
m_{1}$ does not ensures strong-interaction stability of the tetraquark $X_{%
\mathrm{1}}$. It turns out that $X_{\mathrm{1}}$ can decay to pairs of $%
B_{c}^{-}B_{c}^{-}$ and $B_{c}^{\ast -}B_{c}^{\ast -}$ mesons. These
channels form the full width $\Gamma _{\mathrm{1}}=(63\pm 12)~\mathrm{MeV}$
of the state $X_{\mathrm{1}}$ which characterizes it as a "typical"
tetraquark with the relatively modest width. The tetraquark $X_{\mathrm{2}}$
has the mass $m_{2}=(13370\pm 95)~\mathrm{MeV}$ and width $\Gamma _{\mathrm{2%
}}=(79\pm 14)~\mathrm{MeV}$, and is also unstable particle from a family of
doubly charged heavy diquak-antidiquark states.

The tetraquarks $X_{\mathrm{1}}$ and $X_{\mathrm{2}}$ are "pure"
diquark-antidiquark states with fixed internal organizations. In general,
there are other options to model scalar particle $bb\overline{c}\overline{c}$
\cite{Wang:2021taf}. Thus, the scalar tetraquark $bb\overline{c}\overline{c}$
can also have structures $C\gamma _{5}\otimes \gamma _{5}C$, $C\gamma _{\mu
}\gamma _{5}\otimes \gamma ^{\mu }\gamma _{5}C$ and $C\sigma _{\mu \nu
}\otimes \sigma ^{\mu \nu }C$. First two models are composed of color sextet
diquarks, whereas the last one belongs to the triplet representation of the
color group. If exist, scalar tetraquarks $bb\overline{c}\overline{c}$ may
have the same mass and width as aforementioned five basic
diquark-antidiquark states. A physical particle alternatively may be an
admixture of these basic states. But identifications of real scalar
four-quark mesons $bb\overline{c}\overline{c}$ as mixed states, calculations
corresponding mixing parameters are possible only after measuring their
physical parameters.

Nevertheless, even in the lack of such experimental information, we can make
some conclusions about importance of one or the other basic component in a
ground-level scalar particle $bb\overline{c}\overline{c}$. For these
purposes, it is useful to consider an overlap of the currents $J_{1}(x)$ and
$J_{2}(x)$ with the physical state determined by matrix elements $\langle
0|J|X\rangle =\Lambda $%
\begin{eqnarray}
\Lambda _{1} &=&(2.80\pm 0.29)~\mathrm{GeV}^{5},  \notag \\
\Lambda _{2} &=&(1.19\pm 0.14)~\mathrm{GeV}^{5}.
\end{eqnarray}%
Because $\Lambda _{1}>\Lambda _{2}$, the scalar diquark-antidiquark state
couples with a larger strength to the current $J_{1}(x)$ than to $J_{2}(x)$.
In other words, in the ground-level scalar particle, the axial-axial state $%
C\gamma _{\mu }\otimes \gamma ^{\mu }C$ with smaller mass is a dominant
component. Comprehensive analysis of relevant problems requires experimental
data and theoretical analysis of another basic states, which are beyond the
scope of the present article.

As we have mentioned above, tetraquarks $bb\overline{c}\overline{c}$/$cc%
\overline{b}\overline{b}$ were explored in numerous articles using different
methods. The main problem considered in these publications was calculation
of their masses and looking for particles stable against strong decays.
Thus, the mass of scalar states $bb\overline{c}\overline{c}$/$cc\overline{b}%
\overline{b}$ in a color-magnetic-interaction model was estimated within the
limits $12571\div 12597~\mathrm{MeV}$, which is lower than our prediction
for $X_{\mathrm{1}}$ but still higher than the $2B_{c}^{-}$ threshold \cite%
{Wu:2016vtq}.

The scalar tetraquarks $bb\overline{c}\overline{c}$ with both the color
triplet and sextet organizations were studied in Ref. \cite{Wang:2019rdo} in
the framework of nonrelativistic quark models. The mass of $bb\overline{c}%
\overline{c}$ in the model I was found equal to $12863~\mathrm{MeV}$ for $%
\overline{\mathbf{3}_{c}}\otimes \mathbf{3}_{c}$ configuration and to $12850~%
\mathrm{MeV}$ in the case of $\mathbf{6}_{c}\otimes \overline{\mathbf{6}}%
_{c} $ structure. The model II led to slightly higher results: $12915~%
\mathrm{MeV} $ and $12919~\mathrm{MeV}$, respectively. In any case, these
states are above the open heavy-flavor thresholds. Parameters of fully heavy
tetraquarks including $bb\overline{c}\overline{c}$/$cc\overline{b}\overline{b%
}$ ones were also considered in Ref.\ \cite{Liu:2019zuc} in the context of
the potential model by including the linear confining and Coulomb
potentials, as well as spin-spin interactions. It was found that there are
two scalar tetraquarks $bb\overline{c}\overline{c}$ which are admixtures of
the color triplet and sextet states. In the particle with the mass $13039~%
\mathrm{MeV}$ the dominant component is the sextet state, whereas in the
second one with the mass $12947~\mathrm{MeV}$ prevails the triplet
constituent. Both of these states are unstable against strong decays, and
through quark rearrangement can easily dissociate to open-flavor mesons with
appropriate quantum numbers.

Interesting predictions concerning diquark-antidiquark states $bb\overline{c}%
\overline{c}$ were made in Ref.\ \cite{Wang:2021taf}, in which the masses of
the scalar tetraquarks with triplet and sextet compositions were estimated
as $12330_{-150}^{+180}~\mathrm{MeV}$ and $13320_{-240}^{+300}~\mathrm{MeV}$%
, respectively. In accordance with this paper, the state $\overline{\mathbf{3%
}_{c}}\otimes \mathbf{3}_{c}$ is below the $2B_{c}^{-}$ threshold and is
strong-interaction stable tetraquark, whereas the mass of the structure $%
\mathbf{6}_{c}\otimes \overline{\mathbf{6}}_{c}$ overshoots this limit.

As is seen, unstable nature of the scalar tetraquarks $X_{\mathrm{1}}$ and $%
X_{\mathrm{2}}$ was confirmed almost in all publications. Only in Ref.\ \cite%
{Wang:2021taf} the scalar particle $X_{\mathrm{1}}$ (and some other
tetraquarks) was found stable against strong decays. Our prediction for $%
m_{1}$ is considerably higher than $12330_{-150}^{+180}~\mathrm{MeV}$,
whereas\ the result for the mass $m_{2}$ of the state $X_{\mathrm{2}}$
nicely agrees with one from Ref.\ \cite{Wang:2021taf}. Additionally, though
our prediction for the mass splitting $\approx 650~\mathrm{MeV}$ between the
states $X_{\mathrm{1}}$ and $X_{\mathrm{2}}$ is smaller than a gap obtained
there, it is qualitatively consistent with it: In other publications
mentioned above, this splitting is very small.

The widths of the particles $X_{\mathrm{1}}$ and $X_{\mathrm{2}}$ allow us
to interpret them as relatively narrow tetraquarks. The full width of scalar
tetraquark $bb\overline{c}\overline{c}$ was evaluated in Ref.\ \cite%
{Li:2019uch} as well. To this end, the authors used its weak semileptonic
and non-leptonic decays, and estimated the width and lifetime of this
structure as $(3.97\pm 1.50)\times 10^{-9}\ \mathrm{MeV}$ and $(0.17\pm
0.02)\times 10^{-12}\ \mathrm{s}$, respectively. If the tetraquark $X_{%
\mathrm{1}}$ was strong-interaction stable particle, this information would
be very important to search for it in various processes. But, widths $X_{%
\mathrm{1}}$ and $X_{\mathrm{2}}$ amount to a few $\mathrm{MeV}$, therefore
contributions of weak decays to the corresponding parameters are negligible.

There are numerous problems in physics of fully heavy tetraquarks waiting
for their solution. Only relevant experimental measurements combined with
theoretical studies of different four-quark structures can fill up this
segment of exotic hadron spectroscopy.\newline

\section*{ACKNOWLEDGEMENTS}

K. Azizi is thankful to Iran National Science Foundation (INSF) for the
financial support provided under the elites grant number 4025036.

\end{document}